\documentclass[12pt]{article}
\usepackage{graphicx}
\usepackage{amsfonts}
\usepackage{amssymb}
\usepackage{amsthm}
\usepackage{amsmath}
\begin{document}
\begin{titlepage}
\begin{flushright}
hep-ph/0303258{\hskip.5cm}
\end{flushright}
\begin{centering}
\vspace{.3in}
{\bf Gauge Unification and Quark Masses in a Pati-Salam Model from Branes}\\
\vspace{2 cm} { A. Prikas,  N.D. Tracas
} \\
\vspace{1 cm} {\it {
Physics Department, National Technical University,\\
Athens 157 73, Greece}}\\

\end{centering}
\vspace{.1in}

\abstract{\noindent We investigate the phase space of parameters
in the Pati-Salam model derived in the context of D-branes
scenarios, requiring low energy string scale. We find that a
non-supersymmetric version complies with a string scale as low as
$\sim 10$ TeV, while in the  supersymmetric  version the string
scale raises up to  $\sim 2\times 10^7$ TeV. The limited energy
region for RGE running demands a large $\tan\beta$ in order to
have experimentally acceptable masses for the top and bottom
quarks.} \vfill \hrule width 6.7cm
\begin{flushleft}
March 2003
\end{flushleft}

\end{titlepage}

\section{Introduction}


The last few years, there has been considerable work in trying to
derive a low energy theory of fundamental interactions through a
D-brane construction%
\cite{Antoniadis:1990ew,Arkani-Hamed:1998rs,Antoniadis:1998ig,%
Lykken:1996fj,Berkooz:1996km,Balasubramanian:1996uc,Aldazabal:2000sk,%
Antoniadis:2001np,Antoniadis:2000en,Antoniadis:2002qm,Leontaris:2001hh,%
Kokorelis:2002wa,Antoniadis:2002cs}
Recent investigations have shown that there
is a variety of possibilities, concerning the group structure of
the theory as well as the magnitude of the string scale and the
nature of the particle spectrum.

A particularly interesting possibility in this context, is the
case of models with low scale unification of gauge and
gravitational interactions. This is indeed a very appealing
framework for solving the hierarchy problem as one dispenses with
the use of supersymmetry.  There are a number of phenomenological
questions however that should be answered in this case, including
the smallness of neutrino mass\footnote{
For a recent proposal in the context of SM
and the D-brane scenario see%
\cite{Antoniadis:2002qm}}.

Another interesting possibility which could solve a number of
puzzles (as the neutrino mass problem mentioned previously), is
the intermediate scale scenario. A variety of models admit an
intermediate unification scale, however supersymmetry is needed in
this case to solve the hierarchy problem.

In this letter we concentrate on phenomenogical  issues of the
Pati-Salam%
\cite{Pati:1974yy}
gauge symmetry proposed as a D-brane
alternative%
\cite{Leontaris:2001hh}
to the traditional grand
unified version. In particular we investigate the gauge coupling
relations in two cases:
for a non-supersymmetric version and for a supersymmetric one.
In both cases, in order to achieve a low string scale, we relax the
idea of strict gauge coupling unification. However, this should not be considered
as a drawback. Indeed, the various gauge group factors are
associated with different stacks of branes and therefore it is
natural that gauge couplings may differ at the string scale.
In the non-supersymmetric case the string scale could be as small
as a few TeVs. On the other hand, the absence of a large mass
scale puts the see-saw type mechanism (usually responsible for
giving neutrino masses in the experimentally acceptable region) in
trouble. In the supersymmetric case, the string scale is of the
order of $10^3$TeV and a sufficiently suppressed neutrino mass may
be obtained.

\section{The Model}

We assume here a class of  models which incorporate the Pati-Salam
symmetry%
\cite{Pati:1974yy},
having representations that can be
derived within a D-brane construction. In these models, gauge
interactions are described by open strings with ends attached on
various stacks of D-brane configurations and therefore fermions
are constrained to be in representations smaller than the adjoint.
A novelty of these constructions is the appearance of additional
anomalous $U(1)$ factors. At most, one linear combination of these
$U(1)$'s is anomaly free and may remain unbroken in low energies.
As we will see, the role of this extra $U(1)$ is important since
when it is included in the hypercharge definition allows the
possibility of a low string scale.

We start with a brief review of the model%
\cite{Leontaris:2001hh}.
The embedding of the Pati-Salam (PS) model in the brane context
leads to a $SU(4)_C\times SU(2)_L\times SU(2)_R\times U(1)_C\times
U(1)_L\times U(1)_R$ gauge symmetry. Open strings with ends on two
different branes carry quantum numbers of the corresponding
groups. The Standard Model particles appear under the following
multiplets of the PS group:
\begin{equation}\label{quarks}
\begin{split}
F_L&=(\mathbf{4},\mathbf{2},\mathbf{1};1,1,0)\rightarrow Q(\mathbf{3},\mathbf{2},\frac16)
             +L(\mathbf{1},\mathbf{2},-\frac 12)\\
\bar{F}_R&=(\mathbf{\bar{4}},\mathbf{1},\mathbf{2};-1,0,1)\rightarrow
    u^c(\mathbf{\bar{3}},\mathbf{1},-\frac 23)+d^c(\mathbf{\bar{3}},\mathbf{1},\frac 13)+
         e^c(\mathbf{1},\mathbf{1},1)+\nu^c(\mathbf{1},\mathbf{1},0)\\
h&=(\mathbf{1},\mathbf{2},\mathrm{2};0,-1,-1)\rightarrow
      H_u(\mathbf{1},\mathbf{2},\frac 12)+H_d(\mathbf{1},\mathbf{2}-\frac 12)
\end{split}
\end{equation}
where we have also shown the quantum numbers under the three $U(1)$'s \footnote{
 For these assignments see%
\cite{Leontaris:2001hh}}
and the breaking to the SM group. The Higgs which breaks the PS down to the SM is:
\begin{equation}\label{PSHiggs}
\bar{H}=(\mathbf{\bar{4}},1,2;-1,0,\delta)\rightarrow
u_H^c(\mathbf{\bar{3}},\mathbf{1},-\frac 23)+d_H^c(\mathbf{\bar{3}},\mathbf{1},\frac 13)+
         e_H^c(\mathbf{1},\mathbf{1},1)+\nu_H^c(\mathbf{1},\mathbf{1},0)
\end{equation}
The $U(1)$-charge parameter $\delta$ can take two values
$\delta=\pm 1$. Each one of them is associated with a different
symmetry breaking pattern. The down-quark like triplets are the
only remnants after the PS breaking while one Higgs $\bar{H}$ (and
its complex conjugate) is enough to achieve this breaking.
Additional states, such as:
\begin{equation}\label{addit}
\begin{split}
D(\mathbf{6},\mathbf{1},\mathbf{1};2,0,0)&\rightarrow
\tilde{d}^c(\mathbf{\bar{3}},\mathbf{1},\frac 13)+ \tilde{d}(\mathbf{3},\mathbf{1},-\frac 13)\\
\eta(\mathbf{1},\mathbf{1},\mathbf{1};0,0,2)&\\
h_R(\mathbf{1},\mathbf{1},\mathbf{2};0,0,1)&
\end{split}
\end{equation}
can arise which could provide masses to the PS breaking remnants
(colored triplets with down-type quark charges $d_H, d_H^c$) or
break an additional abelian symmetry (by a non-vanishing vev of
$\eta$ and/or $h_R$).

While all three of the $U(1)$'s that come with the PS group are
anomalous, there exists only one combination which is anomaly free
(even from gravitational anomalies):
\begin{equation}\label{U1s}
Y_\mathcal{H}=Y_C-Y_L+Y_R
\end{equation}
where $Y_X,\,X={C,L,R}$ corresponds to the quantum number under the
 $U(1)_X$. None of the SM fermions and Higgs bidoublet
(providing the SM higgses) are charged under this
$U(1)_\mathcal{H}$. To this end, we assume that all anomalous
abelian combinations break and we are left with a gauge symmetry
$SU(4)_C\times SU(2)_L\times SU(2)_R\times U(1)_\mathcal{H}$. The
SM hypercharge is given by the usual PS generators plus a
contribution from the $U(1)_\mathcal{H}$:
\begin{equation}\label{Y}
Y=\frac 12 Y_{B-L}+\frac 12 T_{3R}+cY_\mathcal{H}
\end{equation}
The interesting case is when $c$ differs from zero. Indeed, there
exists a breaking pattern where $c=1/2$ and the parameter $\delta$
determining the $\bar{H}$ charge under the $U(1)_\mathcal{H}$
(namely $\delta-1$), takes the value $\delta=-1$%
\cite{Leontaris:2001hh}.
We are interested in that case and we shall develop the RGE for gauge and
Yukawa couplings running.

\section{Setting the RGE's}

Three different scales appear in our approach: the string scale
$M_U$, the Pati-Salam breaking scale $M_R$ and the low energy
scale $M_Z$. In principle, since the various groups leave in
different stacks of branes, the corresponding gauge couplings may
differ as well. However, in order not to loose predictability at
the unification scale  $M_U$, we require a``petit" unification,
namely $\alpha_4=\alpha_R\neq\alpha_L$ (see%
\cite{Leontaris:2001hh}
for discussion). For
further convenience  we introduce the parameter
$\xi=\alpha_L(M_U)/\alpha_4(M_U)$ ($\alpha_4$, $\alpha_L$ and $\alpha_R$
correspond to the three groups of the model: $SU(4)$, $SU(2)_L$
and $SU(2)_R$). The value of $\alpha_\mathcal{H}$ at $M_U$ is
given by the following relation:
\begin{equation}\label{MU}
\frac{1}{\alpha_\mathcal{H}}=\frac{8}{\alpha_4}+\frac{4}{\alpha_R}
                              +\frac{4}{\alpha_L}
\end{equation}
At $M_R$ we have the following relations due to the Pati-Salam
group breaking:
\begin{equation}\label{MR}
\alpha_3=\alpha_2,\quad\quad
\alpha_2=\alpha_L,\quad\quad
\frac{1}{\alpha_Y}=\frac{2/3}{\alpha_4}+\frac{1}{\alpha_R}+
\frac{c^2}{\alpha_\mathcal{H}}
\end{equation}
where $\alpha_3$, $\alpha_2$ and $\alpha_Y$ correspond to the
three groups of the SM.

As has been mentioned above, the parameter $c$ can take two
acceptable values. The value $c=0$ corresponds to the standard
definition of the hypercharge. Assuming {\it petit unification},
we find%
\cite{Leontaris:2001hh}
$M_U\ge 10^{10}$GeV. The $c=1/2$ introduces a component of
the extra $U(1)_{{\cal H}}$ in $Y$ without affecting the SM charge
assignment. This case allows the possibility of low unification in
the TeV range.
 For the rest of the paper we will work with $c=1/2$. Now
for completeness we give the $\beta$-functions for all groups:
\begin{equation}\label{beta}
\begin{split}
M_U>M>M_R\\
\beta_4&=-\frac{44}{3}+\frac 43 n_g+\frac 13 n_H+\frac 13 n_D\\
\beta_L&=-\frac{22}{3}+\frac 43 n_g+\frac 13 n_h\\
\beta_R&=-\frac{22}{3}+\frac{4}{3}n_g+\frac 13 n_h+
           \frac 23 n_H+\frac 16 n_{h_R}\\
\beta_\mathcal{H}&=\frac{32}{3}n_H+8n_D+\frac 43 n_\eta+\frac 23 n_{h_R}\\
M_Z>M>M_R\\
\beta_3&=-11+\frac 43 n_g+\frac 16 n_{d^c_H}+
        \frac 16(n_{\tilde{d}^c}+n_{\tilde{d}})\\
\beta_2&=-\frac{22}{3}+\frac 43 n_g+\frac 16 (n_{H_u}+ n_{H_d})\\
\beta_Y&=\frac{20}{9} n_g+\frac 19 n_{d^c_H}+\frac 16 (n_{H_u}+ n_{H_d})
                +\frac 19 (n_{\tilde{d}^c}+n_{\tilde{d}})
\end{split}
\end{equation}
where $n_g$ is the number of families ($n_g=3$) while all other notation
is in accordance with that of Eqs.(\ref{quarks}, \ref{PSHiggs}, \ref{addit}).

First we would like to set the range for the parameter $\xi=\alpha_L/\alpha_2$
in order to achieve a low energy $M_U$, while keeping $M_R<M_U$ as an upper
limit and $M_R>1$TeV as a lower limit. We use the following low energy ($M_Z$)
experimental values: $sin^2\theta_W=.23151$, $\alpha_{em}=1/128.9$ and
$\alpha_3=0.119\pm 0.003$. Our particle content is the following:
\begin{equation*}
\begin{split}
&n_g=3, \quad\quad
n_H=1,\quad\quad
n_D=0,\quad\quad
n_h=1,\quad\quad
n_\eta=1,\quad\quad
n_{h_R}=0\\
&n_{H_u}=n_{H_d}=1,\quad\quad
n_{d^c_H}=0\mbox{ or } 1,\quad\quad
n_{\tilde{d}}=n_{\tilde{d}^c}=0
\end{split}
\end{equation*}
and we use one-loop RGE equations.

\begin{figure}[t]
\centering
\includegraphics[scale=.5]{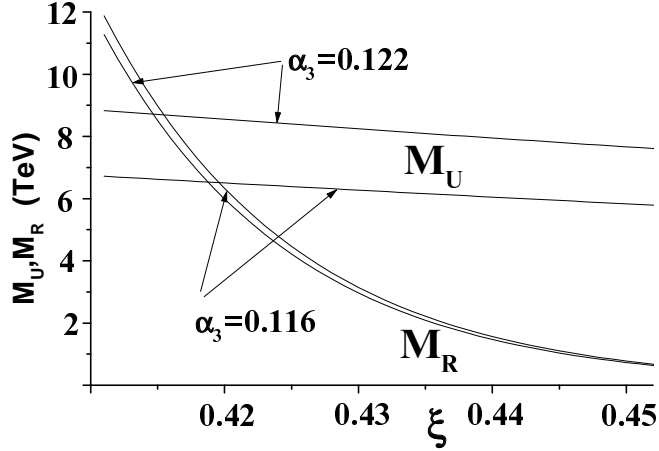}
\caption{The scales $M_U$ and $M_R$ vs the parameter $\xi$. The requirements
1TeV$<M_R<M_U$ sets the range for $\xi$. The particle content has $n_{d^c_H}=0$
(see text).}
\label{MUMRksi}
\end{figure}

In Fig.(\ref{MUMRksi}) we plot $M_U$ and $M_R$ vs $\xi$. The upper
line for $M_U$ and the lower line for $M_R$ correspond to the
highest acceptable value for $\alpha_3$ (with the other lines
corresponding  to the lowest value). The maximum range for
the gauge coupling ratio $\xi$ at $M_U$ is $\xi\sim
(0.413,0.445)$. At the lowest value both scales are of the order
of 9.3TeV while at the highest $M_U\sim 8$TeV. In the case of
absence of non-standard particles, the
region of $\xi$ is $(0.415,0.445)$ and the corresponding values
for the scales are 8.7 TeV and 7.8 TeV. We have also checked that
the gauge couplings stay well in the perturbative region.

We further observe that the $M_R$ and $M_U$  scales merge for the
lower $\xi$ values. Since consistency of the scale hierarchy
demands $M_R\leq M_U$, this implies that there is a lower
acceptable value of $\xi$ or a higher $M_U$ scale as Fig.(1)
shows. On the other hand, experimental bounds on right handed
bosons imply $M_R\gtrsim 1$TeV, this sets the upper bound on $\xi$
or equivalently, the lower bound on $M_U$.

\section{The Supersymmetric Model}

In this section we repeat the above analysis
for the supersymmetric version of the model, where we need the
extra Higgs representation
\begin{equation}\label{PSHiggsSUSY}
H=(\mathbf{4},1,2;1,0,\gamma)\rightarrow
u_H(\mathbf{3},\mathbf{1},\frac 23)+d_H(\mathbf{\bar{3}},\mathbf{1},-\frac 13)+
         e_H(\mathbf{1},\mathbf{1},-1)+\nu_H(\mathbf{1},\mathbf{1},0)
\end{equation}
The charge $\gamma$ is not fully constrained (as opposed to the
case of $\bar{H}$) and, in principle, can take two values
$\gamma=\pm 1$. However, if  supersymmetry is assumed, as the
corresponding charge of the field $\bar H$ has been determined to
$\delta=-1$, the value of $\gamma$ should be fixed to $\gamma=1$.
Further, the following exotic representations could appear
\begin{equation}\label{addit_SUSY}
\begin{split}
\bar{D}(\mathbf{6},\mathbf{1},\mathbf{1};-2,0,0)\\
h_L(\mathbf{1},\mathbf{1},\mathbf{2};0,1,0)\\
\bar{h}_L(\mathbf{1},\mathbf{1},\mathbf{2};0,-1,0)\\
\bar{h}_R(\mathbf{1},\mathbf{1},\mathbf{2};0,0,-1)
\end{split}
\end{equation}

\begin{figure}[!t]
\centering
\includegraphics[scale=.7]{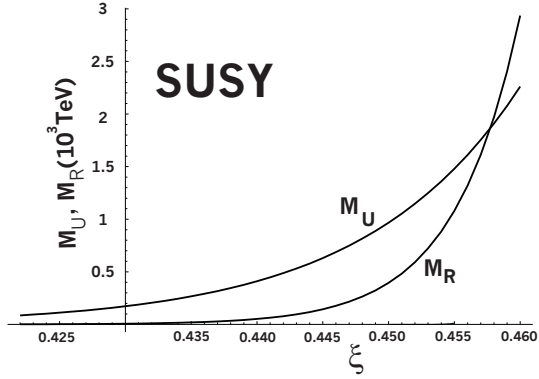}
\caption{The scales $M_U$ and $M_R$ vs the parameter $\xi$  for
the supersymmetric model, ($\gamma=1$, $\delta=-1$). The requirements for
the scales are as in Fig.(1). The
particle content is the minimum one (see text).}
\label{MUMRksi_SUSY}
\end{figure}

\begin{figure}[!b]
\centering
\includegraphics[scale=.7]{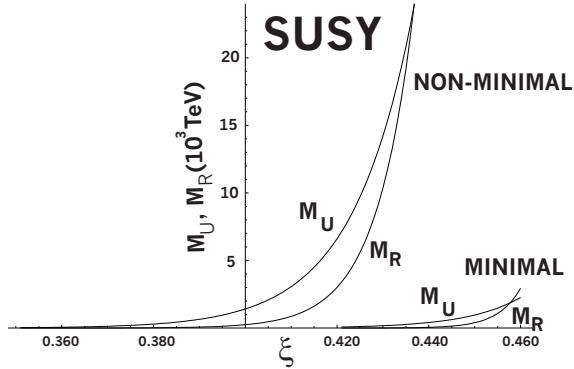}
\caption{The scales $M_U$ and $M_R$ vs the parameter $\xi$ for the
supersymmetric model for the minimal and a non minimal content.}
\label{MUMRksi_SUSY_NONMIN}
\end{figure}

Keeping the same conditions as in the
non-supersymmetric case, Eqs(\ref{MU},\ref{MR}) and
fixing again the value of $c$ to $1/2$, we plot  $M_U$
and $M_R$ vs $\xi$ in Fig.(\ref{MUMRksi_SUSY}). The content is
the minimum possible, i.e.
\begin{equation*}
\begin{split}
     &n_g=3, \quad\quad
      n_H=1,\quad\quad
      n_{\bar{H}}=1,\quad\quad
      n_h=1,\quad\quad
      n_D=0,\quad\quad\\
     & n_\eta=0,\quad\quad
      n_{h_L}=0\quad\quad
      n_{h_R}=0\\
     &n_{H_u}=n_{H_d}=1,\quad\quad
     n_{d^c_H}=0,\quad\quad
     n_{\tilde{d}}=n_{\tilde{d}^c}=0
\end{split}
\end{equation*}

We observe that, in contrast to the
non-supersymmetric case examined in the previous section, here the
limiting case $M_R=M_U$ is realized at the highest $\xi$ value,
while the lower $\xi$ is correlated to the lower acceptable $M_R$
value ($\sim 1$ TeV). The energy scale of $M_U$ and $M_R$ now is
three orders of magnitude higher than the corresponding
non-supersymmetric case.

In Fig.(\ref{MUMRksi_SUSY_NONMIN}) we show the same graph for the
minimal and a non minimal content for the supersymmetric case
($\gamma=1$ and $\delta=-1$). The non minimal content drives the
$\xi$ parameter to lower values but expands the acceptable region
of the scales by almost one order of magnitude.

\section{Yukawa Coupling Running for Top and Bottom}

In the PS model with the minimal Higgs content, the Yukawa
couplings for the top and the bottom quarks are equal at $M_R$,
i.e. $h_t=h_b$. In this section we check whether such a constraint
is compatible with the bottom and top quark masses as they are
measured by the experiments.  If $v_1$ and $v_2$ are the two
v.e.v.'s that correspond to $H_d$ and $H_u$, we have of course:
\[
m_t(m_t)=h_t(m_t)v_2,\quad\quad m_b(m_b)=h_b(m_t)v_1\eta
\]
where the factor $\eta=1.4$ takes care for the QCD renormalisation effects
from the scale $m_t$ down to the mass of the bottom quark.
Since we have two v.e.v.'s (although we do not have supersymmetry), the relation
with $M_Z$ is:
\[
M_Z=\frac 12\sqrt{g^2_2+g_Y^2}\,(v_1^2+v_2^2)=\frac 12\sqrt{g^2_2+g_Y^2}\, v
\]
while we insert, as usual, the parameter $\tan\beta=v_2/v_1$.
The RGE for the two couplings are:
\begin{equation}\label{RGEY}
\begin{split}
16\pi^2\frac{dh_t}{dt}&=h_t\left[\frac 32 h_t^2-\frac 32 h_b^2-
        4\pi\left(\frac{17}{12}\alpha_Y+\frac 94\alpha_2+8\alpha_3\right)\right]\\
16\pi^2\frac{dh_b}{dt}&=h_b\left[\frac 32 h_b^2-\frac 32 h_t^2-
        4\pi\left(\frac{5}{12}\alpha_Y+\frac 94\alpha_2+8\alpha_3\right)\right]\\
\end{split}
\end{equation}
where we have ignored all other Yukawa couplings. We run the
equations from $M_R$ down to scale $M$ where $h_t(M)v_2=M$, which
is the top mass $m_t$.

\begin{figure}[!t]
\centering
\begin{tabular}{cc}
\includegraphics[scale=.35]{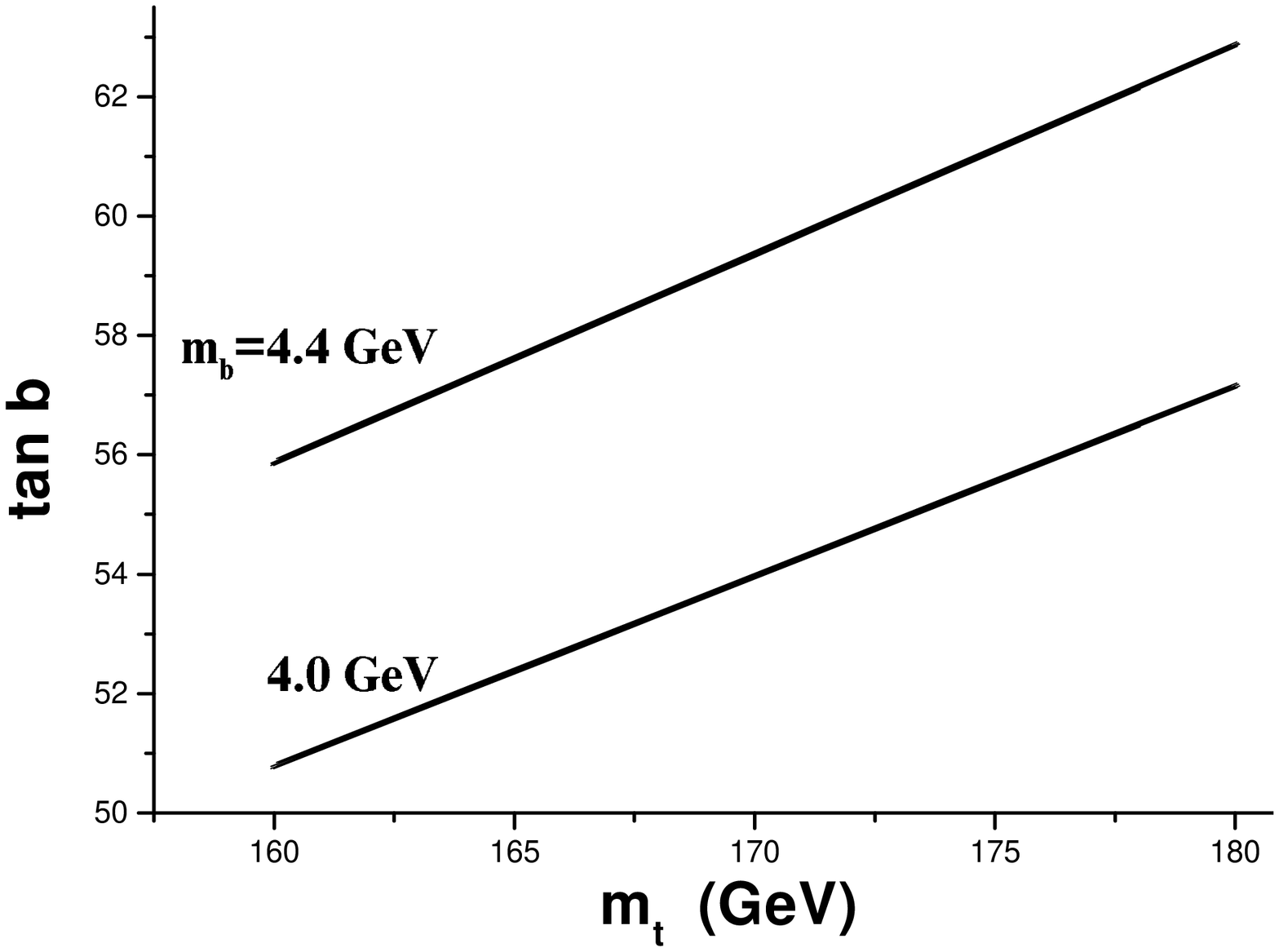}
&
\includegraphics[scale=.45]{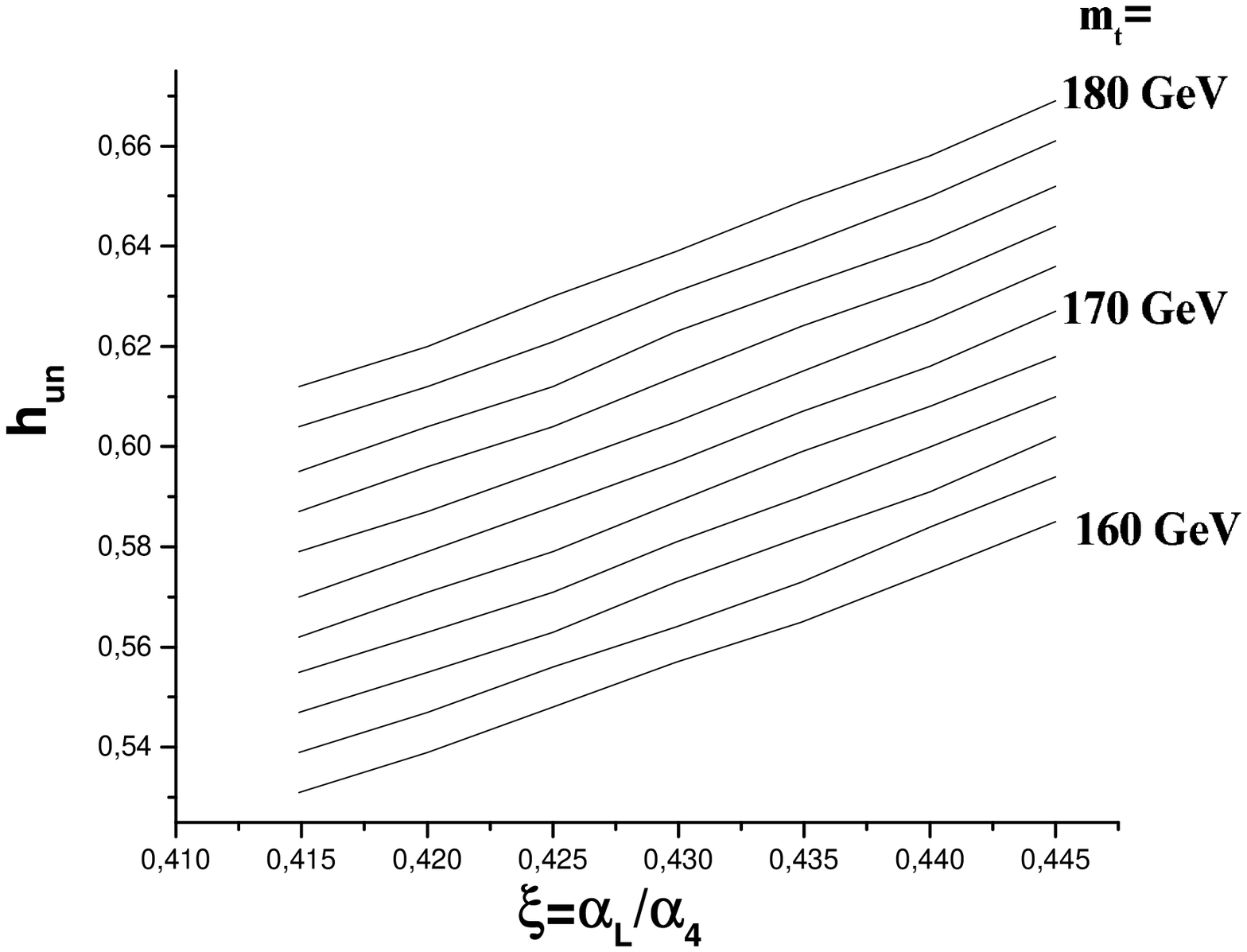}\\
(a)&(b)
\end{tabular}
\caption{(a) The top mass vs $\tan b$ giving $m_b$ in the
experimental range $(4.0-4.4)$GeV and (b) the parameter
$\xi=\alpha_L/\alpha_4$ vs $h(M_R)$ for several values of $m_t$.}
\label{ksi_hun}
\end{figure}

In Fig(\ref{ksi_hun}a) we plot $m_t$ vs $\tan \beta$ in order to
have $m_b$ in the acceptable experimental region (4.0-4.4)GeV. The
choice of $\xi$ (in the acceptable region defined above) makes a
very small effect which shows itself in the thickness of the
lines. Since we require unification of the two Yukawa couplings at
$M_R$, the large difference in the mass of the two quarks can only
be provided by a large angle, therefore the large values of $\tan
\beta$ were expected. Moreover, being in the large $\tan \beta$
regime, $m_t$ changes by a negligible amount as $\tan \beta$
changes to comply with the upper and lower limits of the bottom
mass (remember that $v_2=v\sin b$ while $v_1=v\cos b$).

 The form of the Eq.(\ref{RGEY}) also
shows that the two couplings run almost ``parallel" to each other
and actually the main contribution to the running comes from the
gauge couplings (as we can see in the next figure, the value of
the Yukawas at $M_R$ are small). The corresponding figure with
$n_{d^c_H}=2$ does not show any significant difference.

In Fig.(\ref{ksi_hun}b) we plot the parameter $\xi$ versus the
unified value of the Yukawa coupling at $M_R$, for different
values of $m_t$. The dependence is almost linear with higher value
of $m_t$ requiring higher values of the unified Yukawa coupling
$h$. The absolute value of the Yukawa coupling justifies our
previous claim that the running of $h_b$ and $h_t$ is governed by
the gauge coupling contributions to the RGE equations.

\begin{figure}[!t]
\centering
\begin{tabular}{cc}
\includegraphics[scale=.65]{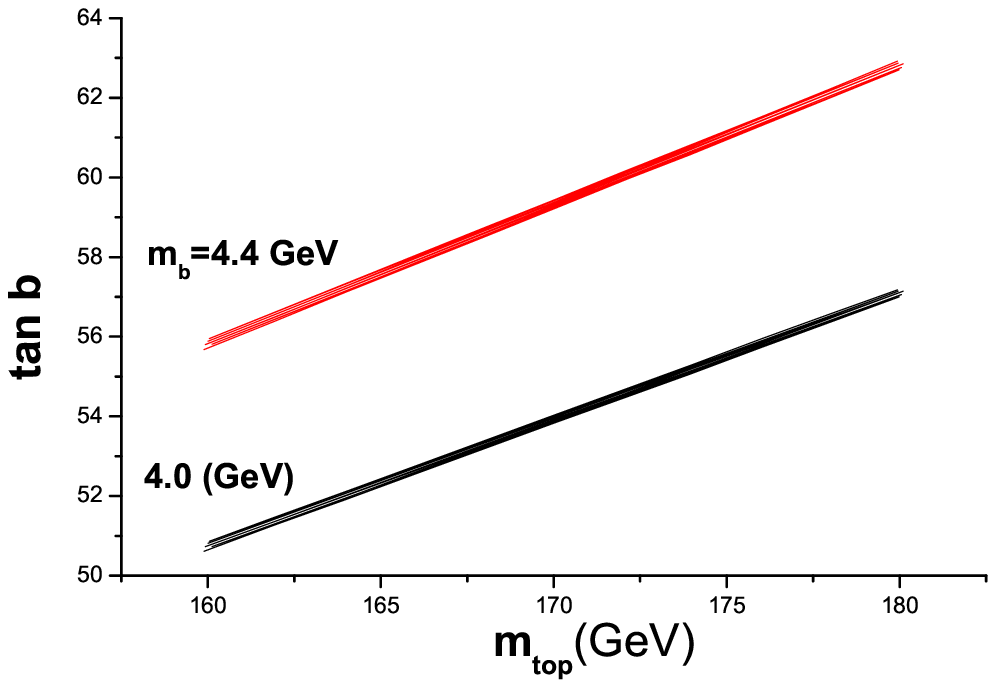}
&
\includegraphics[scale=.75]{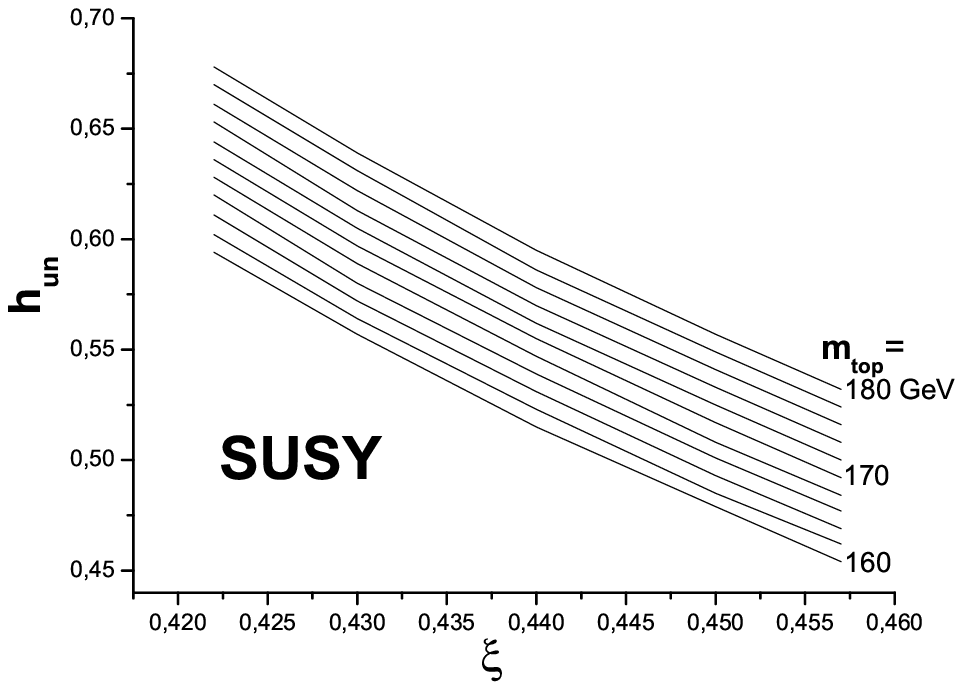}\\
(a)&(b)
\end{tabular}
 \caption{The SUSY case: (a) The top mass vs $\tan \beta$ giving $m_b$
in the experimental range $(4.0-4.4)$GeV
 and (b) The parameter
$\xi=\alpha_L/\alpha_4$ vs $h(M_R)$.}
\label{ksi_hun_SYSY}
\end{figure}

The last figure, Fig.(\ref{ksi_hun_SYSY}), correspond to the supersymmetric case.
The $\tan\beta$ vs $m_t$ figure does not show any significant
difference from the corresponding non-supersymmetric case. On the
contrary, the $h(M_R)$ vs $\xi$ is different. Lower $\xi$ values
corresponds to higher $h(M_R)$ ones while the range of the
acceptable $h(M_R)$ values is a bit broader.

Finally, in figure (\ref{ksi_hun_SYSY}  we plot  the unified value
of the Yukawa coupling $h_u$ versus the $\xi$ parameter. We note
that, -in contrast to the non-supersymmetric case which is
exhibited in figure (5)- here higher $h_u$ values are obtained for
lower ratios $\xi$.

\section{Conclusions}

In the present work, we have examined  the gauge and $bottom-top$
Yukawa coupling evolution in  models with Pati-Salam symmetry
obtained in the context of brane scenarios. In the case of `petit'
unification of gauge couplings, i.e., $a_4=a_R\ne a_L$, it turns
out that in the non-supersymmetric version of the above model one
may have a string scale at a few TeV. Further, assuming $h_b-h_t$
Yukawa unification at the string scale, one finds that the correct
$m_{b,t}$ quark masses are obtained for a $a_4$ approximately
twice as big as $a_L$. A similar analysis for the supersymmetric
case shows that the string scale raises up to $10^7 TeV$ while
$h_b-h_t$ unification reproduces also the right mass relations
$m_b, m_t$.

\vspace{1cm}\noindent
We would like to thank G. Leontaris for valuable discussions and
useful remarks and J. Rizos for helpful discussions.
\hfill

\end{document}